\documentclass[conference, final, letterpaper]{IEEEtran}

\usepackage{hyperref}
\usepackage{ifpdf}
\usepackage{cite}
\ifpdf
    \usepackage[pdftex]{graphicx}
    \graphicspath{{./fig/}}
    \DeclareGraphicsExtensions{.pdf}
\else
    \usepackage[dvips]{graphicx}
    \graphicspath{{./fig/}}
    \DeclareGraphicsExtensions{.eps}
\fi
\usepackage{bbm}
\usepackage{theorem}
\usepackage[cmex10]{amsmath}
\usepackage{amssymb}
\usepackage{stmaryrd}

\newcommand{\prob}{\ensuremath{\mathrm{Pr}}}
\newcommand{\erasure}{\ensuremath{\vartimes}}

\newcommand{\Hd}[1]{\ensuremath{\mathrm{d}_\mathrm{H}(#1)}}

\newcommand{\F}{\ensuremath{\mathbbm{F}}}
\newcommand{\N}{\ensuremath{\mathbbm{N}}}
\renewcommand{\vec}[1]{\ensuremath{\mathbf{#1}}}

\newcommand{\pub}{\ensuremath{\underline{p}}}
\newcommand{\pob}{\ensuremath{\overline{p}}}
\newcommand{\tub}{\ensuremath{\underline{t}}}
\newcommand{\tob}{\ensuremath{\overline{t}}}

\newcommand{\irslambda}{\ensuremath{\frac{\ell+1}{\ell}}}

\newcommand{\T}[2]{\ensuremath{T_{#2}^{(#1)}}}
\newcommand{\Tz}[1]{\ensuremath{\T{z}{#1}}}
\newcommand{\Tzo}{\ensuremath{\T{z}{1}}}
\newcommand{\Tzk}{\ensuremath{\T{z}{k}}}
\newcommand{\Tzz}{\ensuremath{\T{z}{z}}}

\renewcommand{\P}[2]{\ensuremath{P_{#2}^{(#1)}}}
\newcommand{\Pez}{\ensuremath{\P{z}{e, \ell}}}
\newcommand{\Peo}{\ensuremath{\P{1}{e, \ell}}}
\newcommand{\Peinfty}{\ensuremath{\P{\infty}{e, \ell}}}

\newcommand{\pe}{\ensuremath{p_\mathrm{E}}}
\newcommand{\px}{\ensuremath{p_\mathrm{E, X}}}
\newcommand{\De}{\ensuremath{\Delta_\mathrm{E}}}
\newcommand{\Dx}{\ensuremath{\Delta_{\mathrm{E, X}}}}

\newcommand{\estimate}[1]{\tilde{#1}}
\newcommand{\compared}[1]{\widehat{#1}}
\newcommand{\C}{\ensuremath{\mathcal{C}}}
\newcommand{\innerobj}[1]{\ensuremath{#1^\mathrm{i}}}
\newcommand{\Cin}{\innerobj{\C}}
\newcommand{\nin}{\innerobj{n}}
\newcommand{\kin}{\innerobj{k}}
\newcommand{\din}{\innerobj{d}}
\newcommand{\Rin}{\innerobj{R}}
\newcommand{\iin}{\innerobj{\vec{a}}}
\newcommand{\cin}{\innerobj{\vec{c}}}
\newcommand{\rin}{\innerobj{\vec{r}}}
\newcommand{\ein}{\innerobj{\vec{e}}}
\newcommand{\cestin}{\innerobj{\estimate{\vec{c}}}}
\newcommand{\iestin}{\innerobj{\estimate{\vec{a}}}}
\newcommand{\decin}{\innerobj{\mathrm{dec}}}
\newcommand{\outerobj}[1]{\ensuremath{#1^\mathrm{o}}}
\newcommand{\Cout}{\outerobj{\C}}
\newcommand{\nout}{\outerobj{n}}
\newcommand{\kout}{\outerobj{k}}
\newcommand{\dout}{\outerobj{d}}

\newcommand{\iout}{\outerobj{\vec{a}}}
\newcommand{\cout}{\outerobj{\vec{c}}}
\newcommand{\rout}{\outerobj{\vec{r}}}

\newcommand{\cestout}{\outerobj{\estimate{\vec{c}}}}

\newcommand{\rcompout}{\outerobj{\compared{\vec{r}}}}

{\theoremstyle{plain}
   }

{\theoremstyle{plain}
   \newtheorem{theorem}{Theorem}}

{\theoremstyle{plain}
   \newtheorem{lemma}{Lemma}}

{\theoremstyle{plain}
   \newtheorem{corollary}{Corollary}}

\begin{document}

\title{Optimal Thresholds for GMD Decoding with $\irslambda$--extended Bounded Distance Decoders}

\IEEEoverridecommandlockouts

\author{
\authorblockN{Christian Senger, Vladimir R. Sidorenko, Martin Bossert}\thanks{This work has been supported by DFG,
Germany, under grants BO~867/17 and Bo~867/21-1. Vladimir Sidorenko is on leave from IITP, Russian Academy of Sciences, Moscow, Russia.}
\authorblockA{\small Inst. of Telecommunications and Applied Information Theory\\
Ulm University, Ulm, Germany \\
\{christian.senger$\;\vert\;$vladimir.sidorenko$\;\vert\;$martin.bossert\}@uni-ulm.de}
\and
\authorblockN{Victor V. Zyablov}
\authorblockA{\small Inst. for Information Transmission Problems\\
Russian Academy of Sciences, Moscow, Russia \\
zyablov@iitp.ru}
}

\maketitle

\begin{abstract}
We investigate threshold--based multi--trial decoding of concatenated codes with an inner Maximum--Likelihood decoder and an outer error/erasure $\irslambda$--extended Bounded Distance decoder, i.e. a decoder which corrects $\varepsilon$ errors and $\tau$ erasures if $\irslambda\varepsilon+\tau\leq\dout-1$, where $\dout$ is the minimum distance of the outer code and $\ell\in\N\setminus\{0\}$. This is a generalization of Forney's GMD decoding, which was considered only for $\ell=1$, i.e. outer Bounded Minimum Distance decoding. One important example for $\irslambda$--extended Bounded Distance decoders is decoding of $\ell$--Interleaved Reed--Solomon codes. Our main contribution is a threshold location formula, which allows to optimally erase unreliable inner decoding results, for a given number of decoding trials and parameter $\ell$. Thereby, the term {\em optimal} means that the residual codeword error probability of the concatenated code is minimized. We give an estimation of this probability for any number of decoding trials.
\end{abstract}

\section{Introduction}\label{sec:intro}

One of Forney's seminal contributions to algebraic coding was the invention of {\em Generalized Minimum Distance (GMD)} decoding \cite{forney:1966b, forney:1966a}. It provides a means to exploit soft information from the channel using a hard--decision algebraic decoder by multi--trial error/erasure decoding with a varying number of erased unreliable input symbols. Most intriguing about the GMD scheme is that it performs as good as {\em Maximum Likelihood (ML)} decoding if the channel is good. This gives rise to the frequent application of GMD decoding for concatenated codes. There, the inner code is responsible for correcting a considerable amount of transmission channel errors. Thus, the input symbols for the outer decoder can be viewed as being transmitted over a {\em super channel}, which is composed of the transmission channel and the inner decoder. This super channel is always good if the parameters of the inner code are chosen appropriately.

Any decoder's performance can be measured by its guaranteed decoding radius and its residual codeword error probability, the latter one being a function of the channel. The fundamental research problem of threshold--based multi--trial decoding is to find for $z\in\N\setminus\{0\}$ the set of thresholds $\mathcal{T}=\left\{\Tzo, \ldots, \Tzz\right\}$, $\Tzk<\Tz{k+1}$, which optimizes the respective performance measure. Generally, the output of multi--trial decoding is a result list. In this paper, we denote the case that the transmitted codeword is among the elements of the result list as a {\em decoding success}. Both the decoding radius and the residual codeword error probability are to be understood in this context.

Maximization of the guaranteed decoding radius of GMD decoding for concatenated codes was considered by Blokh and Zyablov \cite{blokh_zyablov:1982}. Using Linear Programming, they obtained optimal threshold sets when both inner and outer code are BMD--decoded. In previous work \cite{senger_sidorenko_bossert_zyablov:2008a, senger_sidorenko_bossert_zyablov:2008b}, we generalized their result to the case where the outer code is {\em $\lambda$--extended Bounded Distance ($\lambda$BD)}--decoded for the full (real) range $1<\lambda\leq 2$.

Blokh and Zyablov also considered the probably more practical performance measure, i.e. minimization of the residual codeword error rate. For concatenated codes with inner ML and outer BMD decoding they derived optimal threshold sets using results on the {\em Binary Symmetric Channel (BSC)} error exponent from Gallager \cite{gallager:1965, gallager:1968} and Forney \cite{forney:1968}. In this paper, we tackle the generalization to the case of an outer {\em $\irslambda$--extended Bounded Distance ($\irslambda$BD)} decoder, building up on our previous results \cite{senger_sidorenko_zyablov:2009b}. Thereby, $\ell\in\N\setminus\{0\}$.

The paper is organized as follows. In Section~\ref{sec:conc}, we describe the structure and threshold--based multi--trial decoding of concatenated codes, in Section~\ref{sec:conditions}, we derive necessary and sufficient conditions for an optimal threshold set. We do this on a high level, using the error- and erasure probabilities for each threshold pair $\Tzk$, $\Tz{k+1}$, $k=1, \ldots, z-1$ as parameters. In Section~\ref{sec:exp}, we recall Forney's generalization of Gallager's error exponent of the BSC channel in the error/erasure decoding case.  Simple approximations of the error- and erasure probabilities are derived in Section~\ref{sec:approx}. This allows to analytically calculate the set of optimal thresholds in Section~\ref{sec:thresholds} together with results on the residual codeword error probability. In Section~\ref{sec:conclusions}, we wrap up the paper and draw conclusions for further research.

\section{GMD Decoding of Concatenated Codes}\label{sec:conc}

A concatenated code $\C(n, k, d)$ consists of two constituent codes $\Cin(\F_2; \nin, \kin=m, \din)$ and $\Cout(\F_{2^m}; \nout, \kout, \dout)$. We denote $\Cin$ as the {\em inner} code, $\Cout$ as the {\em outer} code and $\C$ as the {\em concatenated} code. Since $\Cin$ is binary, such is $\C$. W.l.o.g. we restrict ourselves to this most practical case.

An information vector $\iout\in\F_{2^m}^{\kout}$ is first encoded into a codeword $\cout\in\Cout\subseteq\F_{2^m}^{\nout}$. The $2^m$--ary symbols $\outerobj{c}_j$, $j=0, \ldots, \nout-1$, of $\cout$ are then converted into binary vectors $\iin_j\in\F_2^{\kin}$ and encoded into $\cin_j\in\Cin\subseteq\F_2^{\nin}$. The binary matrix consisting of the $\cin_j$ is then transmitted over a BSC channel with crossover probability $e$.

At the receiver, erroneous vectors $\rin_j:=\cin_j+\ein_j$ are received. They are fed into an ML decoder $\decin(\cdot)$ for $\Cin$.The resulting codeword estimates $\cestin_j:=\decin(\rin_j)$ are mapped to their information vectors $\iestin_j$ and converted into symbols $\outerobj{r}_j\in\F_{2^m}\cup\{\erasure\}$, where $\erasure$ is the erasure symbol. The vector $\rout:=(\outerobj{r}_0, \ldots, \outerobj{r}_{\nout-1})$ and the number $z\in\N\setminus\{0\}$ of thresholds are the input for the GMD decoder of $\Cout$.

Inside the GMD decoder, $\rout$ is processed in the following way. First, for every symbol $\outerobj{r}_j\triangleq\iestin_j$, the {\em reliability value} $v_j$,
\begin{equation}\label{eqn:rel}
  v_j:=\left\{\begin{array}{ll}
    \frac{1}{\nin}%
      \ln\left(%
        \frac{\prob\left(\rin_j|\cestin_j\right)}{\sum_{\cin\in\Cin\setminus\left\{\cestin_{\scriptscriptstyle{j}}\right\}} \prob\left(\rin_j|\cin\right)}%
      \right) & ,\,\mathrm{if}\;\cestin_j\in\Cin\\
      0 & ,\,\text{dec. failure}
    \end{array}\right.
\end{equation}
is calculated. Then, the threshold set $\mathcal{T}=\left\{\Tz1, \ldots, \Tzz\right\}$, $\Tzk<\Tz{k+1}$, is applied as
\begin{equation}\label{eqn:erase}
  \outerobj{\compared{r}}_{k, j}:=\left\{\begin{array}{ll}
  \outerobj{r}_j, & \mathrm{if}\;v_j\geq \Tzk,\\
  \erasure, & \mathrm{if}\;v_j<\Tzk
  \end{array}\right.
\end{equation}
 resulting in an {\em input list} $\mathcal{I}:=\left\{\rcompout_1, \ldots, \rcompout_z\right\}$, where $\rcompout_k:=(\outerobj{\compared{r}}_{k, 0}, \ldots, \outerobj{\compared{r}}_{k, \nout-1})$. Thus, a decoding result of the inner decoder is discarded in decoding trial $k$, $k=1, \ldots, z$, if its reliability value falls below the threshold $\Tzk$. Finally, an error/erasure $\lambda$BD decoder (in our case $\lambda=\irslambda$) is applied to every element of $\mathcal{I}$ resulting in a {\em result list} $\mathcal{R}:=\{\cestout_1, \ldots, \cestout_z\}$. Whenever $\cout\in\mathcal{R}$ we have a decoding success.

\section{Necessary and Sufficient Conditions}\label{sec:conditions}

This section generalizes our result from \cite{senger_sidorenko_zyablov:2009b} which was obtained for the simple case where the inner code is BPSK modulation and the outer $\irslambda$BD decoder has parameter $\ell=1$, i.e. it is a BMD decoder.

Considering (\ref{eqn:erase}), it is clear that two cases can occur while advancing from threshold $\Tzk$ to $\Tz{k+1}$. First, it is possible that a correct symbol from the super channel is erased for $\Tz{k+1}$ while it was not for $\Tzk$. Second, it is possible that a wrong symbol is erased for $\Tz{k+1}$ while it was not for $\Tzk$. The probabilities for these two cases are defined by
\begin{align*}
  \pob_k &:= \prob\left(\text{correct symbol erased for}\;\Tz{k+1}\; \text{but not}\; \Tz{k}\right)\\
  \pub_k &:= \prob\left(\text{erroneous symbol erased for}\;\Tz{k+1}\; \text{but not}\; \Tz{k}\right),
\end{align*}
$k=1, \ldots, z-1$. Additionally, we define
\begin{align*}
  p_r &:= \prob(\text{correct symbol never erased})\\
  p_c &:= \prob(\text{correct or erroneous symbol always erased})\\
  p_l &:= \prob(\text{erroneous symbol never erased})
\end{align*}
for the three border cases. We shall find useful approximations for these probabilities in Section~\ref{sec:approx}. Note, that $p_r+p_c+p_l+\sum_{k=1}^{z-1} (\pob_k+\pub_k)=1$. 

With each of the probabilities we associate the number of symbols from $\rout$ falling into the case, i.e. $t_r, t_c, t_l, \tob_k, \tub_k$. Obviously, $t_r+t_c+t_l+\sum_{k=1}^{z-1} (\tob_k+\tub_k)=\nout$.

Let $\varepsilon(k)$ and $\tau(k)$ be the numbers of erroneous and erased symbols, respectively, in $\rcompout_k$. An $\irslambda$BD decoder for $\Cout$ succeeds in decoding $\rcompout_k$ as long as
\begin{equation}\label{eqn:success1}
  \irslambda\cdot \varepsilon(k)+\tau(k) \leq \dout-1.
\end{equation}
This inequality can be expressed by $t_r, t_c, t_l, \tob_k, \tub_k$ as
\begin{equation}\label{eqn:success2}
  \irslambda\left(t_l+\sum_{\nu=k}^{z-1}\tub_\nu\right) + t_c+\sum_{\nu=1}^{k-1} \left(\tob_\nu+\tub_\nu\right) \leq \dout-1,
\end{equation}
since it follows from (\ref{eqn:erase}) and the orderliness of the threshold set that a symbol is erased by $\Tz{k+1}, \ldots, \Tzz$ if it is erased by $\Tzk$. Unequality (\ref{eqn:success2}) is then obtained by simply counting all symbols which are errors and erasures, respectively, for decoding trial $k$ and replacing $\varepsilon(k)$ and $\tau(k)$ in (\ref{eqn:success1}). Let
\begin{equation*}
C_1\hspace{-1mm}:=\hspace{-1.5mm}\left[ \begin{array}{l}
\hspace{-1mm}\forall\, k=1, \ldots, z:\\
\hspace{-1mm}\; \displaystyle\irslambda\left(t_l+\sum_{\nu=k}^{z-1}\tub_\nu\right) + t_c+\sum_{\nu=1}^{k-1} \left(\tob_\nu+\tub_\nu\right) >\dout-1
\end{array}\hspace{-2mm}\right]\hspace{-1.5mm}.
\end{equation*}
Then, by
\begin{equation*}
P_e= \sum_{C} \binom{\nout}{t_l, t_c, t_r, \tub_1, \tob_1, \ldots, \tub_z, \tob_z} 
p_l^{t_l}\, p_c^{t_c}\, p_r^{t_r}
\prod_{k=1}^{z-1} \pub_k^{\tub_k}\, \pob_k^{\tob_k}
\end{equation*}
we obtain an exact formula for the residual codeword error probability of the GMD decoder with $C=C_1$. We can replace the condition by
\begin{equation*}
C_2\hspace{-1mm}:=\hspace{-1.5mm}\left[ \begin{array}{l}
\hspace{-1mm}\forall\, k=1, \ldots, z:\\
\hspace{-1mm}\; \displaystyle\irslambda\left(t_l+\sum_{\nu=k}^{z-1}\tub_\nu\right) + t_c+\sum_{\nu=1}^{k-1} \left(\tob_\nu+\tub_\nu\right) =\dout-1
\end{array}\hspace{-2mm}\right]\hspace{-1.5mm}.
\end{equation*}
to obtain a good approximation of $P_e$ for $C=C_2$. Condition $C_2$ can be compressed to
\begin{equation*}
C_3:=\left[ \irslambda\cdot t_l+t_c+(\ell+1)\sum_{k=1}^{z-1} \tob_k=\dout-1 \right],
\end{equation*}
if we consider $\forall:\,k=1, \ldots, z: \tub_k=\ell\,\tob_k$. The latter set of equalities can be seen by subtracting two subsequent equations of $C_2$ from each other. Since the super channel can be assumed to be good, we can further approximate $P_e$ by
\begin{equation}\label{eqn:Peapprox1}
P_e \approx \max_{C_3}\left\{
p_l^{t_l}\, p_c^{t_c}
\prod_{k=1}^{z-1} {(\pub_k^{\ell}\, \pob_k)}^{\tob_k}
 \right\}.
\end{equation}

The previous observations allow to prove the following theorem.

\begin{theorem}\label{thm:conditions}
The following conditions are necessary and sufficient for an optimal threshold set $\mathcal{T}=\left\{\Tzo, \ldots, \Tzz\right\}$, which minimizes the residual codeword error rate $P_e$.
\begin{align}
  p_l^\frac{\ell}{\ell+1} &= p_c,\label{eqn:condition1}\\
  p_c &= (\pub_1^\ell\, \pob_1)^\frac{1}{\ell+1}\label{eqn:condition2}
\end{align}
and
\begin{equation}\label{eqn:condition3}
  \forall\,k=1, \ldots, z-2: \pub_k^\ell\, \pob_k=\pub_{k+1}^\ell\, \pob_{k+1}.
  \end{equation}
\end{theorem}

\begin{proof}
Equivalent to the maximization in (\ref{eqn:Peapprox1}), we can also express the approximation for $P_e$ in logarithmic form, i.e.
\begin{equation*}
  \ln(P_e)\approx \max_{C_3}\left\{
    t_l\ln(p_l)+t_c\ln(p_c)+\sum_{k=1}^{z-1} \tob_k\ln(\pub_k^\ell\,\pob_k)
  \right\}.
\end{equation*}
Then, the maximization term is a linear function of the $t_l, t_c, \tob_1, \ldots, \tob_z-1$. Thus, its maximum is attained at some boundary point fulfilling condition $C_3$, i.e.
\begin{align*}
\ln(P_e) &\approx \max\left\{
  \frac{\ell\cdot(\dout-1)}{\ell+1}\ln(p_l), (\dout-1)\ln(p_c),\right.\\
  &\quad \hphantom{\max\left\{\right.}%
  \frac{\dout-1}{\ell+1}\ln(\pub_1^\ell\,\pob_1), \ldots\\
  &\quad \hphantom{\max\left\{\right.}%
  \left. \ldots, \frac{\dout-1}{\ell+1}\ln(\pub_{z-1}^\ell\,\pob_{z-1})\right\}.
\end{align*}
In non--logarithmic form:
\begin{align}
P_e &\approx \max\left\{
  p_l^\frac{\ell\cdot(\dout-1)}{\ell+1}, p_c^{\dout-1},\right.\nonumber\\
  &\quad \hphantom{\max\left\{\right.}%
  \left. (\pub_1^\ell\,\pob_1)^\frac{\dout-1}{\ell+1}, \ldots, (\pub_{z-1}^\ell\,\pob_{z-1})^\frac{\dout-1}{\ell+1}\right\}.\label{eqn:Peapprox2}
\end{align}
Let $\mathcal{T}$ fulfill the statement of the theorem and let $\mathcal{T}'$ be a set of thresholds where at least one threshold is different 
 than in $\mathcal{T}$. Assume that $\mathcal{T}'$ is optimal. The only possible 
 way for $\mathcal{T}'$ to achieve a smaller $P_e$ would be to decrease all terms in 
(\ref{eqn:Peapprox2}) simultaneously. This is impossible, decreasing any of the probabilities $p_l, p_c, \pub_1^\ell\,\pob_1, \ldots, \pub_{z-1}^\ell\,\pob_{z-1}$ would increase at least one of the others. This proves that $\mathcal{T}$ is both unique and optimal.
\end{proof}

\section{Forney's Generalization of Gallager's Error Exponent}\label{sec:exp}

Let us for a while consider one specific decoding trial $k$, $k\in\{1, \ldots, z\}$ and the corresponding threshold $T:=\Tzk$. Reliability values are calculated according to (\ref{eqn:rel}) and thresholds are applied as in (\ref{eqn:erase}). Hence, by erasing, the inner ML decoder becomes a decoder with erasing option. Its decoding criterion is defined by
\begin{equation}\label{eqn:criterion}
   \decin(\rin)=\cestin \Longleftrightarrow%
   \frac{\prob(\rin|\cestin)}%
   {\sum_{\cin\in\Cin\setminus\left\{\cestin\right\}} \prob\left(\rin|\cin\right)}%
   \geq \exp T \nin.
\end{equation}
It was shown by Forney that this criterion is optimal in a sense that no other criterion can decrease both the error and error--or--erasure probability \cite{forney:1968}. In the same publication, it was shown that both probabilities can be expressed in terms of Gallager's error exponent for the BSC \cite{gallager:1965, gallager:1968}. They are given by
\begin{align}
  \pe &:= \exp -\left(E_0(\Rin)+s\,T\right)\nin\label{eqn:pe}\\
  \px &:= \exp -\left(E_0(\Rin)-s\,T\right)\nin,\label{eqn:px}
\end{align}
where $E_0(\Rin)$ is Gallager's exponent and $s$ the corresponding optimization parameter, $0< s\leq\frac{1}{2}$.

\section{Approximated Probabilities}\label{sec:approx}

In this section, we shall find simple approximations for the probabilities $\pob_k, \pub_k, p_c$, and $p_l$, which were defined in Section~\ref{sec:conditions}. The approximations are required to obtain an analytic threshold location formula fulfilling the necessary and sufficient conditions of Theorem~\ref{thm:conditions}. Let us start with the following observation. With Gallager's exponent, the error probability of an ML decoder is $\exp -(E_0(\Rin))\nin$. For such decoding, there exists a Hamming distance radius $\Delta\in\N$ such that decoding of $\rin$ with $\Hd{\rin, \cin}\leq\Delta$ almost always succeeds and decoding of $\rin$ with $\Hd{\rin, \cin}>\Delta$ almost always yields an erroneous result. This radius can be thought of as an approximation of the borders of the Voronoi cells of $\cin\in\Cin$. For decoding with erasure option as in (\ref{eqn:criterion}) and the error exponents from (\ref{eqn:pe}) and (\ref{eqn:px}) this gives the following approximations of $\pe$ and $\px$. Recall, that $e$ is the crossover probability of the BSC.
\begin{align*}
  \pe &\approx \sum_{\nu=\De}^{\nin} \binom{\nin}{\nu} e^\nu (1-e)^{\nin-\nu}\\
  \px &\approx \sum_{\nu=\Dx}^{\nin} \binom{\nin}{\nu} e^\nu (1-e)^{\nin-\nu},
\end{align*}
$\De, \Dx\in\N$. Since $\pe\leq \px$, we also have $\De\leq\Dx$. Of course, the probabilities and radii vary for different thresholds. Hence, we append the threshold index as a parameter, i.e. we denote the probabilities and radii for $k=1, \ldots, z$ by $\pe(k), \px(k), \De(k)$, and $\Dx(k)$, respectively. Note, that from  $\Tzk<\Tz{k+1}$ follows $\Dx(k)\geq\Dx(k+1)$ and $\De(k)\leq\De(k+1)$.

Let us consider the probability $p_c$ for a symbol, which is erased for each threshold from $\mathcal{T}$. Its Hamming distance to the transmitted codeword must be at least $\Dx(1)$ and at most $\De(1)-1$. Otherwise, there would be a threshold for which the symbol would not be erased. Consequently, we have
\begin{equation}\label{eqn:pc}
  p_c \approx \hspace{-0.2cm}\sum_{\nu=\Dx(1)}^{\De(1)-1} \binom{\nin}{\nu} e^\nu (1-e)^{\nin-\nu}=\px(1)-\pe(1).
\end{equation}

Now, let us approximate the probability $p_l$ for a wrong symbol which is never erased. Its Hamming distance to the transmitted codeword must be at least $\De(z)$. We obtain
\begin{equation}\label{eqn:pl}
  p_l \approx \sum_{\nu=\De(z)}^{\nin} \binom{\nin}{\nu} e^\nu (1-e)^{\nin-\nu}=\pe(z).
\end{equation}

\begin{figure}[htbp]
\centering
\includegraphics[width=204pt, clip]{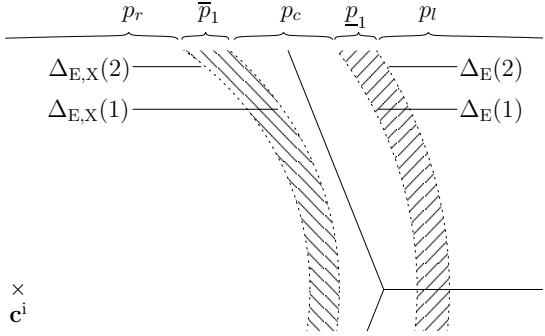}
\caption{Radii for the case $z=2$.}
\label{fig:radii}
\end{figure}

The probabilities $\pob_k, \pub_k$, $k=1, \ldots, z-1$, can be approximated as follows.  Symbols counting towards $\pob_k$ must lie between $\Dx(k+1)$ and $\Dx(k)-1$ as can be seen in Figure~\ref{fig:radii}. For $\pub_k$ we observe that the symbols must lie between $\De(k)$ and $\De(k+1)-1$. We obtain
\begin{align}
  \pob_k &\approx \sum_{k=\Dx(k+1)}^{\Dx(k)-1} \binom{\nin}{\nu} e^\nu (1-e)^{\nin-\nu}\nonumber\\
   &= \px(k+1)-\px(k)-\big(\pe(k)-\pe(k+1)\big)\label{eqn:pob}\\
   \pub_k &\approx \sum_{k=\De(k)}^{\De(k+1)-1} \binom{\nin}{\nu} e^\nu (1-e)^{\nin-\nu}\nonumber\\
   &= \pe(k)-\pe(k+1).\label{eqn:pub}
\end{align}

With equations (\ref{eqn:pc}), (\ref{eqn:pl}), (\ref{eqn:pob}), and (\ref{eqn:pub}) we expressed the probabilities $p_c, p_l, \pob_k$, and $\pub_k$ in terms of the probabilities $\pe(k)$ and $\px(k)$, for which we can use the error- and erasure exponents from (\ref{eqn:pe}) and (\ref{eqn:px}), respectively. After some rather technical simplifications, this yields the following simple expressions.

\begin{lemma}\label{lemma:approx}
The probabilities $p_c, p_l, \pob_k,$ and $\pub_k$ can be approximated by
\begin{align*}
  p_c &\approx \exp -\left(E_0(\Rin)-s\,\Tzo\right)\nin \\
  p_l &\approx \exp -\left(E_0(\Rin)+s\,\Tzz\right)\nin \\
  \pob_k &\approx \exp -\left(E_0(\Rin)-s\,\Tz{k+1}\right)\nin \\
  \pub_k &\approx \exp -\left(E_0(\Rin)+s\,\Tz{k}\right)\nin,
\end{align*}
$k=1, \ldots, z-1$.
\end{lemma}

\section{Optimal Thresholds and Residual Codeword Error Probability}\label{sec:thresholds}

With the results from Sections~\ref{sec:conditions} and \ref{sec:approx} we can now derive an analytic formula for the optimal thresholds $\Tzk$, $k=1, \ldots, z$. Consider Theorem~\ref{thm:conditions}. It basically states a system of $z$ equations, the optimal threshold set with $z$ elements being its solution. Let us express the equations using the approximated probabilities from Lemma~\ref{lemma:approx}. For (\ref{eqn:condition1}), this gives
\begin{align}
  p_l^\frac{\ell}{\ell+1} &= p_c\Longleftrightarrow\nonumber\\
  \frac{\ell}{\ell+1}\left(E_0(\Rin)+s\,\Tzz\right) &= E_0(\Rin)-s\,\Tzo\Longleftrightarrow\nonumber\\
  \frac{E_0(\Rin)}{(\ell+1)s} &= \frac{\ell}{\ell+1}\Tzz+\Tzo.\label{eqn:equation1}
\end{align}
We express (\ref{eqn:condition2}) and (\ref{eqn:condition3}) in the same way and obtain
\begin{align}
  p_c &= (\pub_1^\ell\, \pob_1)^\frac{1}{\ell+1}\Longleftrightarrow\nonumber\\
  \frac{E_0(\Rin)\left(\frac{1-\ell^2}{\ell}\right)}{s} &= \frac{\ell^2+\ell+1}{\ell}\Tzo-\Tz{2}\label{eqn:equation2}
\end{align}
and, $\forall\,k=1, \ldots, z-2$,
\begin{align}
  \pub_k^\ell\, \pob_k &=\pub_{k+1}^\ell\, \pob_{k+1}\Longleftrightarrow\nonumber\\
  0 &= \ell\,\Tzk-(\ell+1)\Tz{k+1}+\Tz{k+2}.\label{eqn:equation3}
\end{align}

\begin{theorem}\label{thm:locationBD}
The optimal threshold set $\mathcal{T}=\left\{\Tzo, \ldots, \Tzz\right\}$ for GMD decoding of a concatenated code, with inner ML and outer $\frac{\ell+1}{\ell}$BD decoding, $\ell\in\N\setminus\{0, 1\}$, is given by
\begin{equation}\label{eqn:TzkBD}
  \Tzk:=\frac{%
    E_0(\Rin)\left(%
      \ell^z(\ell^2+1)-2\ell^k(\ell^2+\ell-1)+\ell^3+\ell^2
    \right)
  }{%
    -s\left(%
      \ell^z(\ell^2+1)+\ell^3-\ell^2-2\ell
    \right)%
  },
\end{equation}
where $E_0(\Rin)$ is Gallager's error exponent for the BSC and $s$ is the corresponding optimization parameter, $0<s\leq\frac{1}{2}$.
\end{theorem}

\begin{proof}
The statement is given by the unique solution of the recurrence relation (\ref{eqn:equation1}), (\ref{eqn:equation2}), and (\ref{eqn:equation3}) for $\ell\in\N\setminus\{0, 1\}$.
\end{proof}

\vspace{-0.28cm}
\begin{corollary}\label{cor:locationBMD}
For outer BMD decoding, i.e. $\ell=1$, the optimal threshold set is given by
\begin{equation*}
  \Tzk:=\frac{E_0(\Rin)(2k-1)}{s(2z+1)}.
\end{equation*}
\end{corollary}

\begin{proof}
The statement is given by the unique solution of the recurrence relation (\ref{eqn:equation1}), (\ref{eqn:equation2}), and (\ref{eqn:equation3}) for $\ell=1$.
\end{proof}

The corollary coincides with the results of Blokh and Zyablov \cite{blokh_zyablov:1982}. Thus, we obtain their result as a special case of our main result, i.e. Theorem~\ref{thm:locationBD}. Note, that both $E_0(\Rin)$ and $s$ -- and thereby also $\Tzk$ -- are functions of the BSC's crossover probability $e$. Also note, that in both cases $\Tzz$ is non--decreasing in $z$.

We shall now state the residual codeword error probability, which can be achieved using an optimal set of thresholds for $\ell>1$. To do this, we return to Theorem~\ref{thm:conditions}, more precisely to (\ref{eqn:Peapprox2}) in its proof. We saw that all terms in the maximization must be equal. Hence, we have the expression
\begin{equation*}
  P_e \approx p_l^\frac{\ell\cdot(\dout-1)}{\ell+1}.
\end{equation*}
Using Lemma~\ref{lemma:approx} gives
\begin{align}
  \Pez &:\approx \left(\exp -\left(E_0(\Rin)+s\,\Tzz\right)\nin\right)^\frac{\ell\cdot(\dout-1)}{\ell+1}\nonumber\\
  &= \exp -\left(\frac{%
  2\ell (\dout-1) (\ell^z-1)
    }{%
      \ell^{z+1}+\ell^{z-1}+\ell^2-\ell-2  
    }\right) E_0(\Rin)\nin,\label{eqn:Pez}
\end{align}
i.e. $\Pez$ is defined by the largest threshold $\Tzz$ within $\mathcal{T}$.

If we are restricted to one single threshold, $\Pez$ becomes
\begin{align}
  \Peo &:\approx \left(\exp -\left(E_0(\Rin)+s\,\T{1}{1}\right)\nin\right)^\frac{\ell\cdot(\dout-1)}{\ell+1}\nonumber\\
  &= \exp -\left(\frac{%
        2\ell (\dout-1) 
      }{%
        2\ell+1
      }\right)E_0(\Rin)\nin.\label{eqn:Peo}
\end{align}

The opposite extremal case is an unlimited number of thresholds. To calculate $\Peinfty$, we require the largest possible threshold, i.e. $\T{\infty}{\infty}$. L'Hospital's rule for (\ref{eqn:TzkBD}) yields
\begin{equation*}
  \T{\infty}{\infty}:=\frac{E_0(\Rin)\left(\ell^2+2\ell-1\right)}{s\left(\ell^2+1\right)}.
\end{equation*}
In the same manner as before, we obtain
\begin{align}
  \Peinfty &:\approx \left(\exp -\left(E_0(\Rin)+s\,\T{\infty}{\infty}\right)\nin\right)^\frac{\ell\cdot(\dout-1)}{\ell+1}\nonumber\\
  &= \exp -\left(\frac{%
        2\ell (\dout-1)
      }{%
        \ell+\frac{1}{\ell}
      }\right)E_0(\Rin)\nin.\label{eqn:Peinfty}
\end{align}


\begin{theorem}\label{thm:Pe}
For GMD decoding of a concatenated code with inner ML and outer $\frac{\ell+1}{\ell}$BD decoding, $\ell\in\N\setminus\{0, 1\}$, and a threshold set $\mathcal{T}=\left\{\Tzo, \ldots, \Tzz\right\}$ from Theorem~\ref{thm:locationBD}, the achievable residual codeword error rate $\Pez$ is in the range
\begin{equation}\label{eqn:boundaries}
  \Peinfty\leq\Pez\leq\Peo,
\end{equation}
where $\Peinfty$ is given by (\ref{eqn:Peinfty}) and $\Peo$ is given by (\ref{eqn:Peo}).
\end{theorem}

For BSC crossover probability $e\longrightarrow 0$, the probabilities in (\ref{eqn:boundaries}) are almost equal. Moreover, $\Peo$ approaches the ML error probability $\exp -E_0(\Rin)\nin$ as $e$ goes to zero. By (\ref{eqn:boundaries}), this happens even faster for $\Pez$ and $\Peinfty$. Morever, our experiments show that the ML error probability is already achieved for moderate channel conditions, especially if $\ell\gg 1$.

\begin{figure}[htbp]
\centering
\includegraphics[width=252pt]{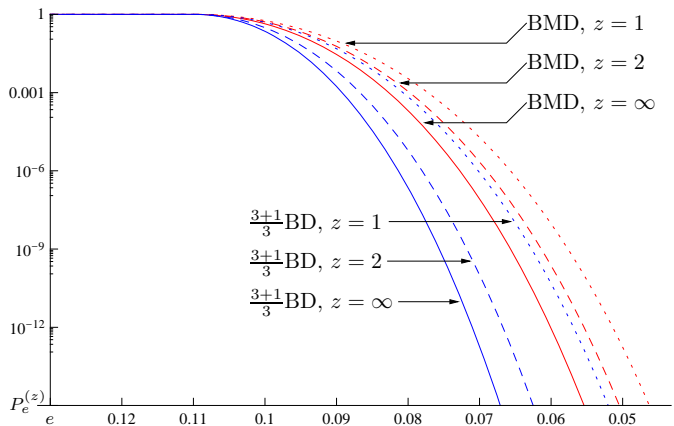}
\caption{Residual codeword error probabilities for $z$ decoding trials.}
\label{fig:errorprob}
\end{figure}

Figure~\ref{fig:errorprob} shows exemplary residual error probability curves for a concatenated code with inner code $\Cin(\F_2; 48, 24, \din)$, outer code $\Cout(\F_{2^{24}}; 255, 223, 33)$ and outer $\frac{3+1}{3}$BD--decoding with a varying number $z$ of optimal thresholds. We observe that $\irslambda$BD decoding always beats BMD decoding. This could have been expected, since it can be shown that
\begin{equation*}
  \P{z}{e, 1}:=\exp -\left(%
    \frac{%
      2z(\dout-1)
      }{%
      2z+1
      }
    \right)E_0(\Rin)\nin\geq \Pez.
\end{equation*}

\section{Conclusions}\label{sec:conclusions}

We investigated threshold--based multi--trial decoding of concatenated codes with inner ML and outer $\irslambda$BD decoding. For any integer number $z$ of decoding trials, i.e. thresholds, we gave an analytic formula for the optimal locations of the thresholds in a sense that the residual codeword error probability is minimized. We showed that for an arbitrary number of thresholds, outer $\irslambda$BD decoding outperforms outer BMD decoding and gave a range of achievable error probabilities. Within this range, the system designer can select $z$ to meet given performance and complexity constraints.

Our results can be applied to standardized concatenated coding schemes, e.g. for the CCSDS Telemetry Channel \cite{ccsds:2002}. It utilizes a set of $\ell$ outer {\em Reed--Solomon (RS)} codes and an inner convolutional code. For a small modification of the standard \cite{schmidt_senger_bossert:2008}, the RS odes can be decoded collaboratively, i.e. they can be considered as an $\ell$--IRS code. For such codes, an efficient error/erasure $\irslambda$BD decoding algorithm has been proposed in \cite{schmidt_sidorenko_bossert:2006c}. Its complexity is the same as for decoding the $\ell$ RS codes separately. Hence, the complexity of multi--trial decoding with the outer $\irslambda$BD decoder grows only linearly in $z$.

Another application of our results is decoding of {\em generalized concatenated codes} \cite{blokh_zyablov:1982, dumer:1998}. There, groups of outer RS codes can be combined into IRS codes as we already pointed out in \cite{senger_sidorenko_bossert_zyablov:2008a}.

\vfill
\def\noopsort#1{}


\begin{thebibliography}{10}

\bibitem{forney:1966b}
G.~D. Forney, ``{G}eneralized {M}inimum {D}istance decoding,'' {\em IEEE Trans.
  Inform. Theory}, vol.~IT-12, pp.~125--131, April 1966.

\bibitem{forney:1966a}
G.~D. {Forney}, {\em Concatenated Codes}.
\newblock Cambridge, MA, USA: M.I.T. Press, 1966.

\bibitem{blokh_zyablov:1982}
E.~L. Blokh and V.~V. Zyablov, {\em Linear Concatenated Codes}.
\newblock Nauka, 1982.
\newblock In Russian.

\bibitem{senger_sidorenko_bossert_zyablov:2008a}
C.~Senger, V.~R. Sidorenko, M.~Bossert, and V.~V. Zyablov, ``Decoding
  generalized concatenated codes using interleaved {R}eed--{S}olomon codes,''
  in {\em Proc. IEEE Int. Symposium on Inform. Theory}, (Toronto, ON, Canada),
  July 2008.

\bibitem{senger_sidorenko_bossert_zyablov:2008b}
C.~Senger, V.~R. Sidorenko, M.~Bossert, and V.~V. Zyablov, ``Multi-trial
  decoding of concatenated codes using fixed thresholds.'' Preprint, 2008.

\bibitem{gallager:1965}
R.~G. Gallager, ``A simple derivation of the coding theorem and some
  applications,'' {\em IEEE Trans. Inform. Theory}, vol.~IT-11, pp.~3--18, Jan
  1965.

\bibitem{gallager:1968}
R.~G. Gallager, {\em Information Theory and Reliable Communication}.
\newblock New York: John Wiley \& Sons, 1968.
\newblock ISBN 0-471-29048-3.

\bibitem{forney:1968}
G.~D. Forney, ``Exponential error bounds for erasure, list, and decision
  feedback schemes,'' {\em IEEE Trans. Inform. Theory}, vol.~IT-14,
  pp.~206--220, March 1968.

\bibitem{senger_sidorenko_zyablov:2009b}
C.~Senger, V.~R. Sidorenko, and V.~V. Zyablov, ``On {G}eneralized {M}inimum
  {D}istance decoding thresholds for the {AWGN} channel,'' in {\em Proc. XII
  Symposium Problems of Redundancy in Information and Control Systems}, (St.
  Petersburg, Russia), May 2009.

\bibitem{ccsds:2002}
Consultative Committee for Space Data Systems, {\em Telemetry Channel Coding},
  October 2002.
\newblock Recommendation for Space Data System Standards, CCSDS 101.0-B-6, Blue
  Book, Issue 6.

\bibitem{schmidt_senger_bossert:2008}
G.~Schmidt, C.~Senger, and M.~Bossert, ``Concatenated code designs with outer
  interleaved {R}eed--{S}olomon codes and inner tailbiting convolutional
  codes,'' in {\em Proc. International ITG Conference on Source and Channel
  Coding}, (Ulm, Germany), January 2008.

\bibitem{schmidt_sidorenko_bossert:2006c}
G.~Schmidt, V.~R. Sidorenko, and M.~Bossert, ``{\noopsort{2}}{C}ollaborative
  decoding of interleaved {Reed--Solomon} codes and concatenated code
  designs,'' {\em IEEE Trans. Inform. Theory}, vol.~IT-55, pp.~2991--3012, July
  2009.

\bibitem{dumer:1998}
I.~I. Dumer, ``Concatenated codes and their multilevel generalizations,'' in
  {\em Handbook of Coding Theory}, vol.~II, ch.~23, Amsterdam: North-Holland,
  1998.
\newblock ISBN 0-444-50087-1.

\end{thebibliography}
\end{document}